\documentclass[aps,prb,groupedaddress,twocolumn]{revtex4}
\usepackage{graphicx}
\usepackage{multirow}

\begin{document}
\title{Magnetic flux disorder and superconductor-insulator transition in nanohole thin films}

\author{Enzo Granato}

\address{Laborat\'orio Associado de Sensores e Materiais,
Instituto Nacional de Pesquisas Espaciais, 12227-010 S\~ao Jos\'e dos
Campos, SP, Brazil}

\begin{abstract}
We study the superconductor-insulator transition in nanohole ultrathin films in a transverse magnetic field
by numerical simulation of a Josephson-junction array model. Geometrical disorder due to the random location of
nanoholes in the film corresponds to random flux in the array model. Monte Carlo simulation in the path-integral
representation is used to determine the critical behavior and the universal resistivity
at the transition as a function of disorder and average number of flux quanta per cell, $f_o$.
The resistivity increases with disorder for noninteger $ f_o$ while it decreases for integer $ f_o$, and
reaches a common constant value in a vortex-glass regime above a critical value of the flux disorder $D_f^c$.
The estimate of $D_f^c$ and the resistivity increase for noninteger $ f_o$  are consistent with recent experiments
on ultrathin superconducting films with positional disordered nanoholes.

\end{abstract}
\pacs{74.81.Fa, 73.43.Nq, 74.40.Kb, 74.25.Uv}

\maketitle


There is growing interest in the superconductor-insulator (SI) transition in
ultra-thin films with a lattice of nanoholes
\cite{valles07,valles08,batur11,kopnov,valles15,valles16}. This system is an important testing ground for models of the universality class of the quantum phase transition since the patterned nanostructure provides a sensitive probe for distinguishing between phase and amplitude fluctuations of the superconducting order parameter.
The magnetoresistance oscillatory behavior at low magnetic fields near the transition is analogous to the one
observed in microfabricated  Josephson-junction arrays, which undergo a SI transition
due to the small electrical capacitance of the superconducting grains
\cite{fazio,sondhi,geerligs,vdzant,gk90}. This common feature results from
phase coherence effects, which can be described by the same generic model of phase fluctuations of the superconducting order parameter, a Josephson-junction array model, with a wider applicability. In fact, it is  closely related to the Bose-Hubbard model, where Cooper pairs interact on a lattice potential, in the limit of a large number of bosons per site \cite{sondhi,cha3}, to the quantum rotor model \cite{cha2,cha3,Kope} and to ultracold atoms on optical lattices \cite{polak,nature,micnas}. For a periodic nanohole film at low magnetic fields, the simplest model consists of a frustrated array of superconducting "grains", where the phase is well defined locally, coupled by Josephson junctions or weak links on a periodic lattice, with the lattice of nanoholes corresponding to the dual lattice, which acts as a vortex pinning center \cite{eg13,eg16}. The number of flux quanta per unit cell of the nanohole lattice, which is proportional to
the external magnetic field, corresponds to the frustration parameter $f$ of the Josephson-junction array model.
The zero-temperature quantum phase transition in the array model, driven by the competition between the charging energy and Josephson-coupling energy at different frustration parameters, corresponds to the SI transition in the nanohole film in the external magnetic field. The resistivity at the transition is expected to be finite and universal \cite{fisher,cha2,cha3,fisher2}, depending only on the universality class of the transition, which generally changes in the presence of a magnetic field and disorder.

Very recently, intriguing experimental results have been obtained near the SI transition in thin films with a disordered triangular lattice of nanoholes with controlled amount of positional disorder \cite{valles15,valles16}.
Such disorder leads to spatial variations in the magnetic flux per unit cell, which increases with the
magnetic field, similar to the effects of geometrical disorder in  microfabricated Josephson-junction arrays \cite{gk86,benz88}. Magnetoresistance oscillations decrease in amplitude and disappear above a critical value of flux disorder. However, the resistivity at successive field-induced transitions {\it increases} with flux disorder, in apparent disagreement with predictions of universality \cite{fisher,cha2,cha3} and a previous numerical simulation \cite{stroud08}, which show a {\it decrease} of the resistivity.

In this work, we study the SI transition in geometrically disordered nanohole thin films
by numerical simulation of a Josephson-junction array model with flux disorder. Geometrical disorder due to the random locations of the nanoholes in the film corresponds to random flux in the array model. Monte Carlo (MC) simulation in the path integral representation is used to determine the critical behavior and the resistivity at the transition as a function of flux-disorder strength $D_f$ and average number of flux quanta per cell, $ f_o$. It is found that the resistivity at the transition increases with disorder for noninteger $ f_o$ while it decreases for integer $ f_o$, and  reaches an approximately common constant value in a vortex-glass regime above a critical value $D_f^c$. The distinct behavior for noninteger $f_o$ results from the interplay of vortex-lattice commensurability  and flux-disorder effects. The estimate of $D_f^c$ and the resistivity increase for noninteger $ f_o$  are in good agreement with available experimental data  on positional disordered nanohole thin films \cite{valles15} for noninteger $ f_o$ while it calls for further measurements for integer $f_o$.


We consider a Josephson-junction array model, which allows for both
flux disorder and charging effects \cite{gk86,stroud08,fazio}, described by the Hamiltonian
\begin{equation}
{\cal H} = -{{E_c}\over 2} \sum_i n_i^2 - \sum_{<ij>} E_{ij} \cos ( \theta_i -
\theta_{j}-A^o_{ij}- t_{ij}). \label{hamilt}
\end{equation}
The first term in Eq. (\ref{hamilt}) describes quantum
fluctuations induced by the charging energy, $ E_c  n_i ^2/ 2$,  of a
non-neutral superconducting grain located at site $i$ of a periodic reference lattice, where $E_c= 4 e^2/C$,
$e$ is the electronic charge, and
$n_i= -i \partial /\partial \theta_i $ is the operator, canonically conjugate to the phase operator $\theta_i$, representing the deviation of the number of Cooper pairs from
a constant integer value. The effective capacitance to the ground of each grain $C$
is assumed to be spatially uniform, for simplicity.
The second term in (\ref{hamilt}) is the
Josephson-junction coupling between nearest-neighbor grains
described by phase variables  $\theta_i$.
The effect of the magnetic field $\bf B$ applied in the perpendicular ($\hat z$-direction)
appears through the link variables $A^o_{ij}$ and $t_{ij}$, which satisfy the constraints $\sum_{ij} A^o _{ij} = 2 \pi f_o$ and $\sum_{ij} t_{ij} = 2 \pi  \delta f_p$, where the gauge-invariant sums $\sum_{ij}$  are over the links $ij$ surrounding the site $p$ of the plaquette centers. $f_o$ is a uniform constant parameter and $\delta f_p$ is a spatially varying random variable with zero average. The effects of the positional disorder of the nanoholes, which corresponds to random plaquette areas  $S_p$ of the array, can be  incorporated in this model by identifying $f_o$ as the average number of flux quanta per plaquette $BS_o/\Phi_o$, where  $\Phi_o=hc/2e$ is the flux quantum, and $S_o$ as the uniform plaquette area of the  reference lattice. $\delta f_p$  then represents the  additional random flux  $ f_o \delta S_p/S_o $, where $\delta S_p = S_p -S_o$. Previous work on the SI transition \cite{stroud08} studied this model defined on a square lattice for integer $f_o$ and uncorrelated disorder in $t_{ij}$.  In order to compare with available experimental data for superconducting films with a triangular lattice of nanholes in the weak disorder limit
\cite{valles07,valles15}, we consider here the array model defined on a honeycomb lattice \cite{note0} and take $\delta f_p$ as an uncorrelated random variable. For convenience, we use a uniform disorder distribution  $\delta f_p  = D_f[-1,1]$, with the random-flux disorder strength $D_f= f_o D_a$, where $D_a$ measures the disorder in the areas $\delta S_p/S_o$.
Experimentally, the flux disorder $D_f$ can be varied by changing $f_o$ via the external field or the geometrical disorder $D_a$ using different samples \cite{valles15}.
We also allow for bond disorder in the form of random  Josephson couplings \cite{note}  $E_{ij}= E_J \ e_{ij}$, where $e_{ij}= 1 \pm D_b$ with equal probability and disorder parameter strength  $D_b$. In the numerical simulations described below we set $D_b=0.3$ but its value does not change the main results.  With this choice the  magnetoresistance  behavior of films with a triangular lattice of nanoholes without flux disorder \cite{valles07,kopnov} can already be described by the array model \cite{eg13,eg16}. Here we consider the effects of increasing  the flux disorder  $D_f$  for integer $f_o=n$ and noninteger rational values $f_o=n+1/q$ of the frustration parameter.

To study the quantum phase transition at zero temperature, we employ the imaginary-time path-integral formulation of the model \cite{sondhi}. In this representation, the two-dimensional (2D) quantum model of Eq. (\ref{hamilt}) maps into a (2+1)D classical statistical mechanics problem. The extra dimension corresponds to the
imaginary-time direction.
The classical reduced Hamiltonian can be written as
\begin{eqnarray}
H= &&-\frac{1}{g} [ \sum_{\tau,i}
\cos(\theta_{\tau,i}-\theta_{\tau+1,i}) \cr &&
+\sum_{<ij>,\tau} e_{ij}\cos(\theta_{\tau,i}-\theta_{\tau,j}-A^o_{ij}-t_{ij}) ],
\label{chamilt}
\end{eqnarray}
where $ e_{ij}=E_{ij}/ E_J$ and $\tau$ labels the sites in the discrete time direction.
The ratio $g =(E_c/E_J)^{1/2}$, which drives the SI transition for the
model of Eq. (\ref{hamilt}), corresponds to an effective "temperature" in the
3D classical model of Eq. (\ref{chamilt}).
In general, a quantum phase transition shows intrinsic anisotropic scaling, with  different diverging correlation
lengths $\xi$ and $\xi_\tau$ in the spatial and imaginary-time directions
\cite{sondhi}, respectively, related by the dynamic critical exponent $z$
as $\xi_\tau \propto \xi^z$.
The classical Hamiltonian of Eq. (\ref{chamilt}) can be viewed as an
XY model on a layered honeycomb lattice, where  frustration effects exist
only in the honeycomb layers. Randomness in $e_{ij}$ and $t_{ij}$ corresponds to disorder
completely correlated in the time direction.
The honeycomb lattice is defined on a rectangular geometry with linear size given by a dimensionless length $L$.
In terms of $L$, the linear size in the $\hat x$  and $\hat y$ directions
correspond to $L_x=L\sqrt{3}/2$ and $L_y=\frac{3}{2}L$, respectively. We choose a gauge where
$A_{ij}= 2 \pi f n_y$, on alternating (tilted) bonds along the rows in the
$\hat x$ direction numbered by the integer $n_y$ and $A_{ij}=0$ otherwise.

Equilibrium MC simulations for $E_c > 0$ are carried out using the
3D classical Hamiltonian in Eq. (\ref{chamilt}) regarding $g$ as a "temperature"-like parameter.
The parallel tempering method \cite{nemoto} is used in the simulations
with periodic boundary conditions, as in previous work \cite{eg16}.
The finite-size scaling analysis is performed for different sizes $L$ with the constraint
$ L_\tau =a L^z $, where $a$ is a constant aspect ratio. This choice simplifies the scaling analysis, otherwise
an additional scaling variable $L_\tau/L^z$ would be required to describe the scaling functions.
The value of $a$ is chosen to minimize the deviations of $ a L^z$ from integer numbers. However, this requires one to know the value of the dynamic exponent $z$ in advance.
Since the exact value of $z$ is not known, we follow a two-step approach.
First, we obtain an estimate of $g_c$ and $z$ from  simulations performed with a {\it driven} MC dynamics method, which has been used in the context of the 3D XY-spin glass model \cite{eg04}. Then, these initial estimates are improved by finding the best data collapse for the finite-size behavior of the phase stiffness in the time direction $\gamma_\tau$, obtained by the equilibrium MC method. For the driven MC method, the layered honeycomb model of Eq. (\ref{chamilt}) is viewed as a 3D superconductor and the corresponding "current-voltage" scaling near the transition is used to determine the critical coupling and critical exponents \cite{wengel}. In the presence of an external driving perturbation $J_x$ ("current density") which couples to the phase difference $\theta_{\tau,i + \hat x}-\theta_{\tau,i}$ along the $\hat x$ direction, the classical Hamiltonian of Eq. \ref{chamilt} is modified to
\begin{eqnarray}
H_J= H -\sum_{i,\tau} J_x (\theta_{\tau,i+\hat x}-\theta_{\tau,i}).
\label{driven}
\end{eqnarray}
When $J_x \ne 0$,  the system is out of equilibrium since the  total energy is unbounded.
The lower-energy minima occur at phase differences $\theta_{\tau,i+\hat x}-\theta_{\tau,i}$,
which increase with time $t$, leading to a net phase slippage
rate proportional to $ V_x = <d(\theta_{\tau,i+\hat x}-\theta_{\tau,i})/dt > $, corresponding to the average
"voltage" per unit length. The MC simulations are carried out using the Metropolis algorithm and
the time dependence is obtained by identifying the time $t$ as the MC time. The measurable quantity
of interest is the phase slippage response ("nonlinear resistivity") defined as $R_x = V_x/J_x$.  Similarly, we define $R_\tau$ as the phase slippage response to the applied perturbation $J_\tau$ in the layered (imaginary-time) direction.
Above the phase-coherence
transition, $g > g_c$, $R_x$ should approach a nonzero value when $J_x \rightarrow 0$ while it should  approach zero below the transition. From the nonlinear scaling behavior near the transition of a sufficiently large system, one can extract the critical coupling $g_c$, and the critical exponents $\nu$ and $z$.
In the absence of charging effects, $R_x$ remains zero below a critical value $J_x=J_c$, which provides an
estimate of the critical current for  the model of Eq. (\ref{hamilt}), when $E_c=0$.

\begin{figure}
\includegraphics[bb= 1.5cm 0.5cm  12.cm  8.7cm, width=7.5cm]{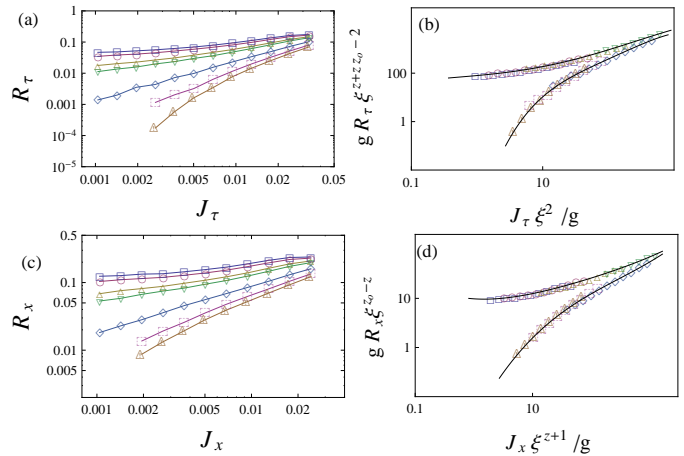}
\caption{ Phase slippage response in (a) the imaginary-time direction  $R_\tau$  and (c) spatial  direction $R_x$  for $\bar f= n + 1/6$, near the transition. Flux-disorder strength $D_f=0.7$  and system size  $L=60$.
The couplings $g$ from top down are $1.48$, $1.47$, $1.45$, $1.44$, $1.41$, $1.39$, $1.38$.
(b) and (d): scaling plots corresponding to (a) and (c), respectively, for data near the transition with $\xi=|g/g_c-1|^{-\nu}$ and the same parameters  $g_c=1.426 $, $z_o=2.3$, $z=1.2$ and $\nu=1.1$. }
\label{grhotau}
\end{figure}

We show in detail the results for $f_o = n + 1/6$ and $D_f=0.7$. This
value of frustration was chosen to allow a comparison with the available experimental data \cite{valles15}. Fig. \ref{grhotau}  shows the behavior of the nonlinear phase slippage response $R_x$ and $R_\tau$ as a function of the applied perturbation $J_x$ and $J_\tau$, respectively, for different couplings $g$ and  large system size.
The behavior is consistent with a phase-coherence transition at an apparent
critical coupling in the range $g_c \sim 1.41 - 1.44$. For $g > g_c$, both
$R_x$ and $R_\tau$ tend to a finite value while for $g < g_c$, they
extrapolate to low values.
The critical coupling $g_c$ and critical exponents $\nu$ and $z$ can then be obtained from the
best data collapse satisfying the scaling behavior close to the transition.
The required scaling theory is described in detail in ref. \onlinecite{girvin}.
$R_x$  and $R_\tau$ should satisfy the scaling forms
\begin{eqnarray}
g R_x \xi^{z_0-z} & = & F_\pm (J_x \xi^{z+1}/g),  \cr
g R_\tau \xi^{z + z_0 z - 2} & = & H_\pm (J_\tau \xi^2/g),
\label{Rxy}
\end{eqnarray}
where $z_o$ is an additional critical exponent describing the MC relaxation times, $t_{x}\sim \xi^{z_o}$ and  $t_{\tau}\sim \xi_\tau^{z_o}$, in the spatial and imaginary-time directions, respectively,  and  $\xi=|g/g_c-1|^{-\nu}$.
The + and -  signs correspond to $g  > g_c$ and $g  < g_c$, respectively. The two scaling forms are the same when $z=1$, corresponding to isotropic scaling.  The joint scaling plots according to Eqs. \ref{Rxy} are shown in  Fig. \ref{grhotau}, obtained by adjusting the unknown parameters, providing the estimates $g_c=1.426 $, $z_o=2.3 $, $z=1.2 $ and $\nu=1.1$.

\begin{figure}
\includegraphics[bb= 1.5cm 0.5cm  14.cm  10cm, width=7.5cm]{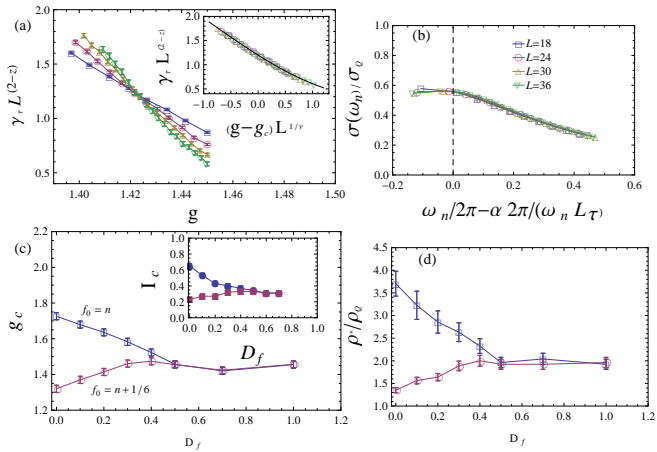}
\caption{ (a) Phase stiffness in the imaginary time direction $\gamma_\tau $ for
different system sizes $L$, near the transition point estimated from Figs. \ref{grhotau}. $L_\tau=a L^z$,
with aspect ratio $a=0.628$ and $z=1.25$.
Inset: scaling plot of $\gamma_\tau $ with $g_c =1.424 $ and $ \nu =0.97 $ .
(b) Scaling plot of conductivity $\sigma(iw_n)$ at the critical coupling $g_c$ with
$\alpha=0.15$. The universal conductivity  is given by the intercept with the $x=0$
dashed line, leading to  $\frac{\sigma^*}{\sigma_Q}= 0.56(3)$.
(c): Critical coupling $g_c$ at different values of the average frustration $f_o$ and increasing flux-disorder strength $D_f$.
Inset: behavior of the corresponding critical currents $I_c$ at $E_c=0$.
(d): Resistivity $\rho^*=1/\sigma^*$ in units of $\rho_Q=1/\sigma_Q$ at the transition for the different average frustrations indicated in (c) and increasing flux disorder $D_f$.  }

\label{stifftau}
\end{figure}

The above estimate  of $g_c$ and $z$ does not take into account the finite-size effects. It assumes that the system is sufficient large and the coupling is not too close to $g_c$ such that the correlation length is smaller than the system size. To improve these estimates we consider the finite-size behavior of the phase stiffness in the imaginary time direction $\gamma_\tau$.
The phase stiffness $\gamma_\tau$, which is
a measure of the free energy cost to impose an infinitesimal phase twist in the time
direction, is given by \cite{cha2}
\begin{eqnarray}
\gamma_\tau=\frac{1}{L^3 g^2}[g <\epsilon_\tau>
- < I_\tau^2 > + < I_\tau >^2]_D,
\label{eqstiff}
\end{eqnarray}
where $\epsilon_\tau = \sum_{\tau,i} \cos(\theta_{\tau,i} -\theta_{\tau+1,i})$ and   $I_\tau= \sum_{\tau,i}\sin(\theta_{\tau,i} -\theta_{\tau+1,i})$. In Eq. (\ref{eqstiff}), $< \ldots>$ represents a MC average
for a fixed disorder configuration and $[  \ldots ]_D$ represents an average over different disorder configurations.
In the superconducting phase $\gamma_\tau$ should be finite, reflecting the existence of phase coherence, while in the
insulating phase it should vanish in the thermodynamic limit.
For a continuous phase transition, $\gamma_\tau$ should satisfy the finite-size scaling form
\begin{equation}
\gamma_\tau L^{2-z} = F(L^{1/\nu} \delta g),
\label{rhotau}
\end{equation}
where $F(x)$ is a scaling function and $\delta g = g-g_c$. This scaling form implies that data for $ \gamma_\tau L^{2-z} $ as a function of $g$, for different system sizes $L$, should cross at the critical coupling $g_c$. Fig. \ref{stifftau}a shows this crossing behavior obtained near the initial estimate of $g_c$ by varying $z$ slightly from its initial value. In the Inset of this Figure, we show a scaling plot of the data according to the scaling form of Eq. \ref{rhotau}, which provides the final estimates $g_c= 1.424 $  and $\nu = 0.97$.

We have also determined the universal conductivity at the critical point from the frequency and finite-size dependence of the phase stiffness $\gamma(w)$ in the spatial direcion, following the scaling method described by Cha {et al.} \cite{cha2,cha3}.
The conductivity is given by the Kubo formula
\begin{equation}
\sigma = 2 \pi \sigma_Q \lim_{w_n\rightarrow 0}  \frac{\gamma(i w_n)}{w_n},
\end{equation}
where $\sigma_Q=(2 e)^2/h$ is the quantum of conductance and $\gamma(i w_n)$ is a frequency dependent phase stiffness
evaluated at the finite frequency $w_n=2 \pi n /L_\tau$, with $n$ an integer. The phase stiffness in
the $\hat x$ direction is given by
\begin{eqnarray}
\gamma= C [g < \epsilon_x >
-< |I(i w_n)|^2 >
+ < |I(i w_n)| >^2]_D,
\end{eqnarray}
where $C= 1/((4/3 \sqrt{3})  N L_\tau g^2 )$,  $N$ is the total number of sites in each layer,
\begin{eqnarray}
\epsilon_x&=&\sum_{\tau,j} (\hat x \cdot \hat u_{j,j+\hat x })^2 e_{i,j+\hat x}\cos(\Delta_x \theta_{\tau,j}) ,\cr
I(i w_n)&=&\sum_{\tau,j} (\hat x \cdot \hat u_{j,j+\hat x }) e_{i,j+\hat x} \sin(\Delta_x \theta_{\tau,j}) e^{i w_n \tau},
\end{eqnarray}
$\hat u_{j,j+\hat x}$  is a unit vector between nearest neighbors sites and $\Delta_x \theta_{\tau,j}= \theta_{\tau,j}-\theta_{\tau,j+\hat x}- A^o_{j,j+\hat x} - t_{j,j+\hat x}$.
At the transition, $\gamma(i w_n)$ vanishes linearly with frequency and $\sigma$ assumes a universal value  $\sigma^*$, which can be  extracted from its
frequency and finite-size dependence \cite{cha2}
\begin{equation}
\frac{\sigma(iw_n)}{\sigma_Q} = \frac{\sigma*}{\sigma_Q}
- c (\frac{w_n}{2 \pi} - \alpha \frac{2 \pi}{w_n L_\tau}) \cdots
\label{cond}
\end{equation}
The parameter $\alpha$ is determined from  the best data collapse of the frequency
dependent curves for  different systems sizes  in a plot of $\frac{\sigma(iw_n)}{\sigma_Q}$ versus
$x=(\frac{w_n}{2\pi} - \alpha \frac{2\pi}{w_n L_\tau})$. The universal conductivity is
obtained from the intercept of these curves with the line $x=0$.
The calculations were performed for different system sizes with $L_\tau=a L^z$, using the above estimates of $z$ and $g_c$. From the scaling behavior in Fig. \ref{stifftau}b we obtain $\sigma^*/\sigma_Q = 0.56(3)$, where the estimated uncertainly is mainly the result of the error in the coupling $g_c$.

We have performed extensive calculations as a function of the flux disorder strength $D_f$ for integer  $ f_o = n$ and  noninteger $ f_o = n + 1/6$. The behavior of the critical couplings $g_c$ for the SI transition as a function of $D_f$ is shown in Fig. \ref{stifftau}c and  the corresponding behavior of the resistivity at the transition $ \rho^* = 1/ \sigma^*$ is shown in Fig. \ref{stifftau}d. Disorder changes significantly the values of the critical coupling and resistivity for small $D_f$ while they remain essentially unchanged and frustration independent above a critical value  $D^c_f \sim 0.5 $.
Below $D^c_f$, the resistivity at the transition increases with disorder for noninteger $f_o$ but it decreases for integer $ f_o$. This critical disorder $D^c_f$ should correspond to
a transition into a vortex glass regime, where one expects that $g_c$ should be insensitive to the value of the frustration. 
Similar behavior is also expected for the critical current in absence of charging effects \cite{gd01}. Calculations for the critical current for the model of Eq. \ref{hamilt} with $E_c=0$ using the driven MC dynamics are shown in the Inset of Fig. \ref{stifftau}c. The transition from a low-disorder regime, where the critical current is sensitive to frustration, to a glassy regime occurs at approximately the same critical value $D^c_f$.

The results for noninteger $ f_o$ are in good agreement with available experimental observations on ultrathin superconducting films with positional disordered nanoholes \cite{valles15}. As in other calculations of the resistivity at the transition \cite{cha3,cha2,gk90,micnas,eg16}, the obtained value  differs significantly from the experimental value. However, the trend as a function of disorder and the magnetic field dependence should be consistent with experiments. In fact, the resistivity for large flux disorder found experimentally for the field-induced SI transition in the nanohole films \cite{valles15} is a factor of $1.8(2)$ higher than in the absence of disorder, which agrees reasonably well with our numerical estimate of $1.5(3)$ for noninteger $f_o$ in Fig. \ref{stifftau}d.
The experimental data also allows a rough estimate of the critical exponent product $z \nu \sim 1.4(4) $, from the expected scaling behavior of the resistivity derivative  at the transition \cite{fisher2} as a function of temperature $T$,  $\partial \rho/\partial B \propto T^{- 1/(z \nu) }$. Our numerical estimate  $z \nu = 1.21(5)$ is compatible with the experimental value although the errorbars are large. Moreover, the critical disorder strength below which magnetoresistance oscillations are observed experimentally \cite{valles15}, $\delta f_c \sim 0.3$, can also be compared with the critical disorder strength $D^c_f\sim 0.5 $ found numerically.
These oscillations occur below $D_f^c$, where the critical coupling for the SI transition  $g_c$ in Fig. \ref{stifftau}c is sensitive to frustration, with decreasing amplitude as the flux disorder $D_f=f_o D_g$  approaches $D_f^c$ for increasing frustration.
Since in the present calculations $\delta f$ is uniformly distributed, rather than approximately Gaussian distributed as in the experiments, a conversion factor is required for comparing the critical values. Requiring the variance of both distributions to be the same leads to an equivalent flux disorder strength $ \sim 0.5/\sqrt{3}=0.29$, which is in reasonable agreement with the experimental value. For integer $f_o$, the resistivity in Fig. \ref{stifftau}d for large flux disorder decreases by a factor of $1.7(3)$. A much larger decrease has been found previously \cite{stroud08} for the model of Eq. \ref{hamilt} defined on a square lattice with uncorrelated disorder in $t_{ij}$. Unfortunately, experimental data for integer $f_o$, including $f_o=0$, on the same sample are not available yet to make a comparison to the numerical results. However, the resistivity found in recent experiments for larger flux disorder \cite{valles16} decreases by a factor $ \sim 2$ compared with earlier measurements on samples without flux disorder \cite{valles07}, which is compatible with the present calculations.

The change of the resistivity and the different behavior for noninteger $f_o$ as a function of $D_f$, can be understood as the interplay of vortex-lattice commensurability  and flux disorder effects. In absence of disorder, the SI transition for noninteger $f_o$  is in a different universality class from the zero field case \cite{gk90}. The net circulating currents around each plaquette, introduced by the external field, correspond to a pinned commensurate vortex lattice which changes the ground-state symmetry. Since the resistivity  depends on the universality class \cite{fisher}, its value for noninteger $f_o$ can be significantly different. For $f_o=1/2$ on a square lattice \cite{gk90,vdzant,cha2,micnas}, for example,  it decreases by a factor of $2$. In the present case of a honeycomb lattice, the SI transition for $f_o=1/2$ is yet in another universality class different from the square lattice \cite{eg16}, and the resistivity decreases by a factor of approximately $4$. On the other hand, for large flux disorder, where there is a vortex glass phase for both integer and noninteger $f_o$, the universality and the resistivity should be the same, since the vortices are in a highly disordered configuration.


In conclusion, we found that the resistivity at the SI transition increases with magnetic-flux disorder $D_f$ for noninteger frustration $f_o$ while it decreases for integer $f_o$, and  reaches an approximately common value in a vortex-glass regime for $D_f > D_f^c$.
In the simplest scenario, one expects different critical behavior for weak  and  strong disorder.
Although the obtained constant value of the resistivity for  $D_f > D_f^c$ indicates universal behavior in a different universality class, the variation of the resistivity for small disorder, however, may be a result of crossover effects due the limited system sizes. In the experiments, temperatures not sufficiently low should have similar effects.
In the absence of such effects, the results can not rule out a truly non universal behavior.
The results could also be tested experimentally in microfabricated Josephson-junction arrays with controlled parameters. However, for a more realistic description of these systems,  disorder from offset charges and dissipation effects \cite{fazio}, which have been neglected in the present model, should be taken into account.

\medskip

The author thanks J. M. Valles Jr.  for helpful discussions and suggestions.
This work was supported by  S\~ao Paulo Research Foundation (FAPESP, Grant \#
2014/15372-3) and computer facilities from CENAPAD-SP.


\begin{thebibliography}{9}
\bibitem{note0}
The charging energy $E_c$ of such an array, associated with a node of the nanohole film,
was estimated to be too large \cite{valles07,valles08}. However, in the present model,
it is an effective parameter, which  incorporates other effects leading to $E_c$
comparable to $E_J$.
\bibitem{note} Bond disorder washes out the secondary minimum at $f=1/3$ predicted by
the model for the magnetoresistance oscillations \cite{eg13,eg16} when $D_f=0$, leading to a behavior consistent with experiments \cite{valles07,kopnov}.
It is also generated by the random nanohole positions without coupling to the external field.
\end{thebibliography}
 \end{document}